# A Catalog of Prominence Eruptions Detected Automatically in the SDO/AIA 304 Å Images

S. Yashiro[1,2*], N. Gopalswamy[2], S. Akiyama[1,2], and P.A. Mäkelä [1,2]

[1]The Catholic University of America, Washington, DC 20064, USA

[2]NASA Goddard Space Flight Center, Greenbelt, MD 20771, USA

*Corresponding author, email: seiji.yashiro@nasa.gov

**Abstract:**

We report on a statistical study of prominence eruptions (PEs) using a catalog of these events routinely imaged by the Atmospheric Imaging Assembly (AIA) on board the Solar Dynamics Observatory (SDO) in the 304 Å pass band. Using an algorithm developed as part of an LWS project, we have detected PEs in 304 Å synoptic images with 2-min cadence since May 2010. A catalog of these PEs is made available online (https://cdaw.gsfc.nasa.gov/CME_list/autope/). The 304 Å images are polar-transformed and divided by a background map (pixels with minimum intensity during one day) to get the ratio maps above the limb. The prominence regions are defined as pixels with a ratio $\geq 2$. Two prominence regions with more than 50% of pixels overlapping are considered the same prominence. If the height of a prominence increases monotonically in 5 successive images, it is considered eruptive. All the PEs seen above the limb are detected by the routine, but only PEs with width ≥15° are included in the catalog to eliminate polar jets and other small-scale mass motions. The identifications are also cross-checked with the PEs identified in Nobeyama Radioheliograph images (http://solar.nro.nao.ac.jp/norh/html/prominence/). The catalog gives the date, time, central position angle, latitude, and width of the eruptive prominence. The catalog also provides links to JavaScript movies that combine SDO/AIA images with GOES soft X-ray data to identify the associated flares, and with SOHO/LASCO C2 images to identify the associated coronal mass ejections. We examined the statistical properties of the PEs and found that the high-latitude PE speed decreased with the decreasing of the average polar magnetic field strength of the previous cycle.

## 1. Introduction

Prominence eruptions (PEs) are important for a comprehensive understanding of coronal mass ejections (CMEs) because they are closely related (see e.g., Gopalswamy 2015). PE and CME catalogs have also enabled important research on the solar source regions and the long-term behavior of solar magnetic regions (e.g., Gopalswamy et al. 2003, 2012; Shimojo, 2013; Yashiro et al. 2004). The Nobeyama Radio Heliograph (NoRH) has been detecting PEs routinely since 1992 (Shimojo et al., 2006), but the data coverage is limited (~8 hours/day) because it is a ground-based instrument. Another catalog that compiled PEs in STEREO/SECCHI 304 Å images using a tool known as Solar Limb Prominence Catcher and Tracker (SLIPCAT2; Wang et al. 2010) is also available online, but only for observations from 2007 to 2012. Here we present a catalog of PEs detected semi-automatically using the 304 Å images obtained by the Atmospheric Imaging Assembly (AIA, Lemen et al. 2012) on board the Solar Dynamics Observatory (SDO). The catalog distinguishes regular PEs and surges. In addition to a list of PEs, their flare and CME associations, we present a statistical study of these prominences.

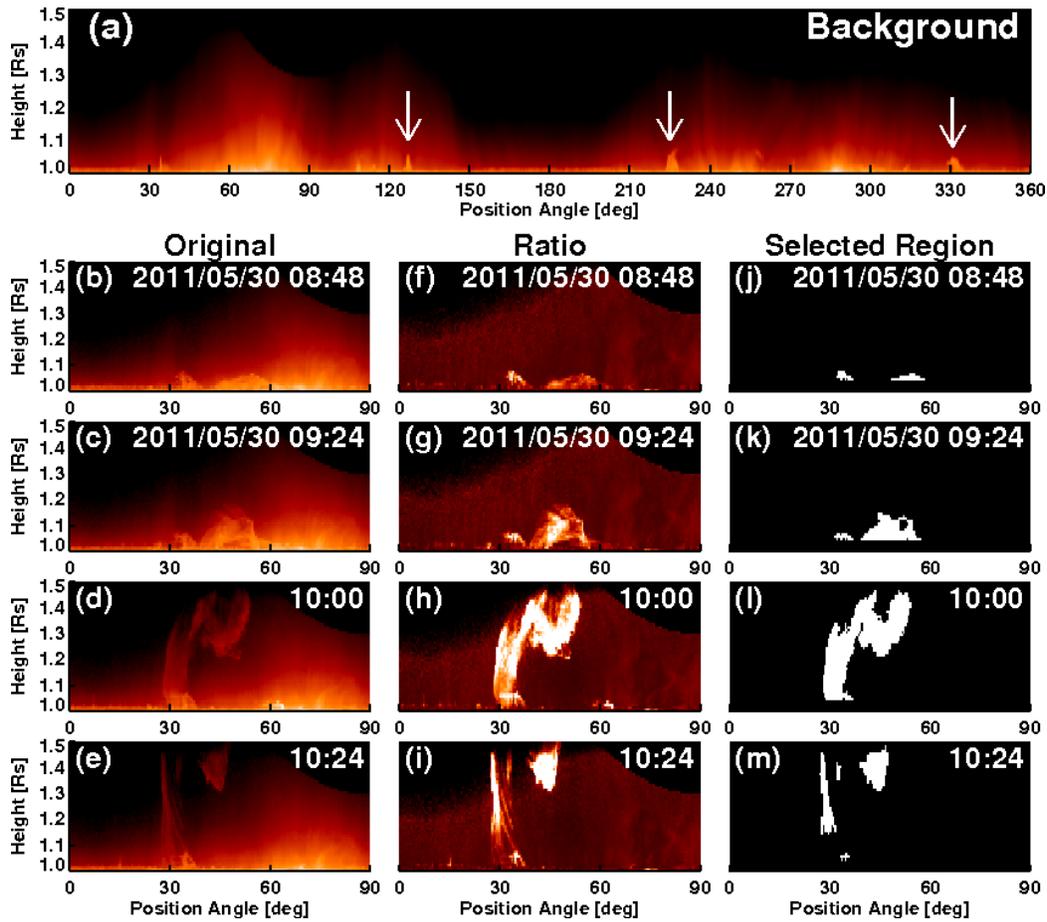

*Figure 1 – An example on how our algorithm detects a prominence eruption that appeared above the NE limb around 9:00 UT on 2011 May 30. All data are transformed into the polar coordinates. The position angle (X-axis) is measured counterclockwise from north in degrees. (a) A polar-transformed background image for data between 9 UT to 10 UT. Different background images are used for different hours. (b-e) The prominence eruption seen in the original images. (f-i) The prominence eruption seen in the ratio images. (j-m) The prominence eruption region identified by the algorithm. See text for details.*

## 2. Data and Method

The SDO/AIA instrument has been obtaining EUV images at 304 Å every 12 seconds with a spatial resolution of 0.6 arcsec and an image size of 4096x4096 pixels. Since erupting prominences are large-scale structures and typical eruptions lasts for more than several tens of minutes, the synoptic data (1024x1024 pixels, at 2 min cadence) provided by the Joint Science Operations

Center (JSOC) at Stanford University are good enough to detect prominence eruptions. The AIA synoptic data have been flat-fielded, dark subtracted, de-spiked, corrected for bad pixels, and aligned image top with the solar north (level 1.5). The spatial resolution has been reduced to 2.4 arcsec by binning. For easy handling of the outward motion of eruptions, the 304 Å images are transformed to the polar coordinates (*PA*, *H*), where *PA* is the position angle measured counterclockwise from north in degrees and *H* is the heliocentric distance in solar radii (Rs). The dimensions of the transformed data are 720x50 pixels corresponding the PA and radial distance ranges of 0°–360° and 1.0–1.5 Rs with resolutions of 0.5° and 0.01 Rs, respectively. For saving computer resources, the data on the solar disk are discarded.

The next step is to obtain the background images in order to enhance the signal levels of erupting prominences. The 304 Å channel has been used to observe the chromospheric materials by the He II resonance line, but the Si XI line emission from corona becomes dominant above the limb (Labrosse & McGlinchey, 2012). The background intensity for a pixel is defined as the minimum intensity of the pixel during a period of 24 hours. Bright transient features are filtered out and the stationary structures remain as the background. Since brightness of the corona keeps changing, we derive one background image per hour, i.e., 24 background images per day. Figure 1a shows the background image for the data between 9 UT to 10 UT on 2011 May 30. This background image was derived using data between 21:30 UT on 2011 May 29 and 21:30 UT on 2011 May 30. The X-axis is the position angle: 0° (360°), 90°, 180°, and 270° correspond to north, east, south, and west, respectively. The background intensity is bright above active regions and dark above the polar regions. Small bright features at *PA* of 125°, 225°, and 330° (pointed by arrows) are prominences that were quiet for more than 24 hours.

Panels (b-e) in Figure 1 show the northeast portion (*PA* 0°–90°) of the polar-transformed data at 8:48 UT, 9:24 UT, 10:00 UT, and 10:24 UT on 2011 May 30. A prominence started to rise slowly at around 9:00 UT and reached the edge of the AIA field of view (FOV) at 10:12 UT. The upward motion of the prominence can be seen, but the prominence brightness is similar to that of the active region. In order to see the prominence motion more clearly, the transformed data are divided by the corresponding background image (different background images are used for different times of observation as noted above). The ratio images are shown in Panels (f-i). Normally the brightness of the eruptive prominence decrease with time: the prominence in panel (d) is darker than that in panel (c). However, as the prominence moves to higher height where the background is darker, the contrast in the ratio maps increases as can be seen in the panels (g) and (h). Because the stationary bright corona at low latitudes ($PA$ $60°$–$90°$) is not seen in the ratio map, the eruption can be traced more easily. We define prominences as regions of pixels having a ratio of 2 or greater. Increasing of the threshold (e.g., 2.5 or greater) will not reduce the number of bright events, will reduce the number of faint events, and will reduce the number of false detections. The detection score might be better with the higher threshold but detected eruption speed will de degraded because the higher threshold can catch the bright inner part only. The best threshold could be different for different prominences, but this threshold looks reasonable for most prominences. Panels (j-m) show binary maps of the selected eruptive prominence regions. We should note that this method can take "eruptive" prominences only. As we see in panel (a), the stationary prominences are considered as a part of the background. Prominences are detected only when they move.

The cross calibration of AIA 304 A channel with SDO/EVE and Hinode/EIS showed that the effective area of AIA 304 degraded (Boerner et al. 2014). Generally speaking, the instrument degradation affects the automated detection capability. However, we could not find noticeable degradation of the detection capability because we have used the ratio maps. The degradation affects both PE and background brightness, so we don't need to adjust the threshold.

Occasionally, independent eruptions would occur simultaneously at the different locations. In order to separate these independent eruptions, we used the "LABEL_REGION" function in IDL to perform the connected-component analysis (or blob coloring) for each selected eruption region to assign a unique index. Two regions connected at a pixel's vertex are labeled by different indexes.

Pixels at the north (left and right edge pixels in Polar-Coordinate array) are treated as neighborhood. In the successive images, if more than 50% pixels of a selected region overlap with a selected region in the previous image, both regions are considered to be parts of the same eruption.

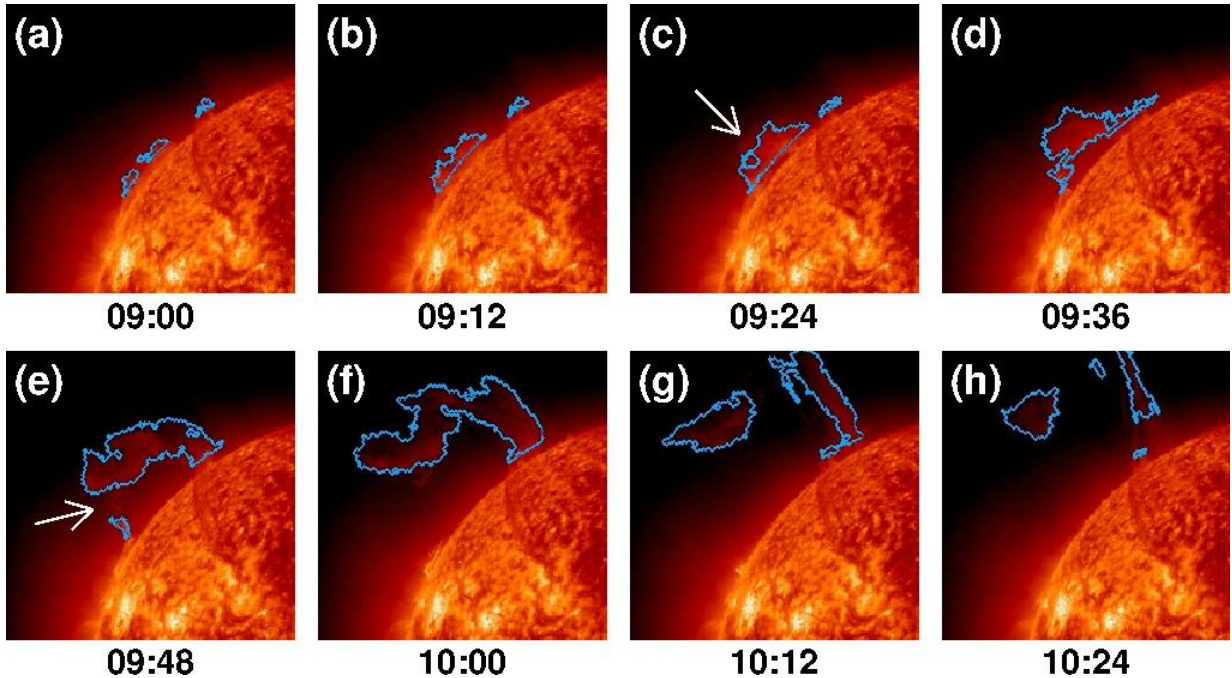

*Figure 2 – (a-h) The same prominence eruption as in Fig. 1. The blue contours enclose the selected regions of the eruptive prominence. White arrows point the prominence regions where do not meet the ratio threshold.*

The selected eruption regions in the polar coordinates are mapped on the original observed data to verify the algorithm results. The blue contours in Figure 2 indicate the prominence boundary determined by the algorithm. Some parts of the prominence are not selected (e.g., in e and f), but the overall structure of the prominence is well identified.

In Fig. 2c, the leading edge of the prominence is not properly tracked because it does not meet the threshold ratio at some locations. The prominence eruption on 2011 May 30 in Fig. 2 is well defined, but identifying the leading edge can be difficult sometimes. On the other hand, the centroid of the prominence is stable because the position is determined by all pixels identified as the prominence. Irregular changes of the boundary can be smoothed, so tracking the centroid may be preferred for robust prominence detection.

We consider a detected event as an eruption if its centroid-height increases monotonically in 5 successive images (4-step increases). Many false detections due to fluctuation of the coronal intensity (i.e., noise) are filtered out by this criterion. It is possible that some true eruptive events with short duration (~ 10 minutes) could also be filtered out. This is a limitation of this algorithm. However, this is not a significant problem because such short-duration prominence eruptions are very rare. Changing the criterion of monotonic increase from 4steps to 6steps reduces the number of false detections but increases the number of missing true eruptions. The true-false validation of this algorithm is discussed in Section 3.2.

*Figure 3. SDO/AIA PE list available on line at the CDAW Data Center. The items in blue have links to movies that show the PEs in association with GOES lightcurves and SOHO/LASCO images.*

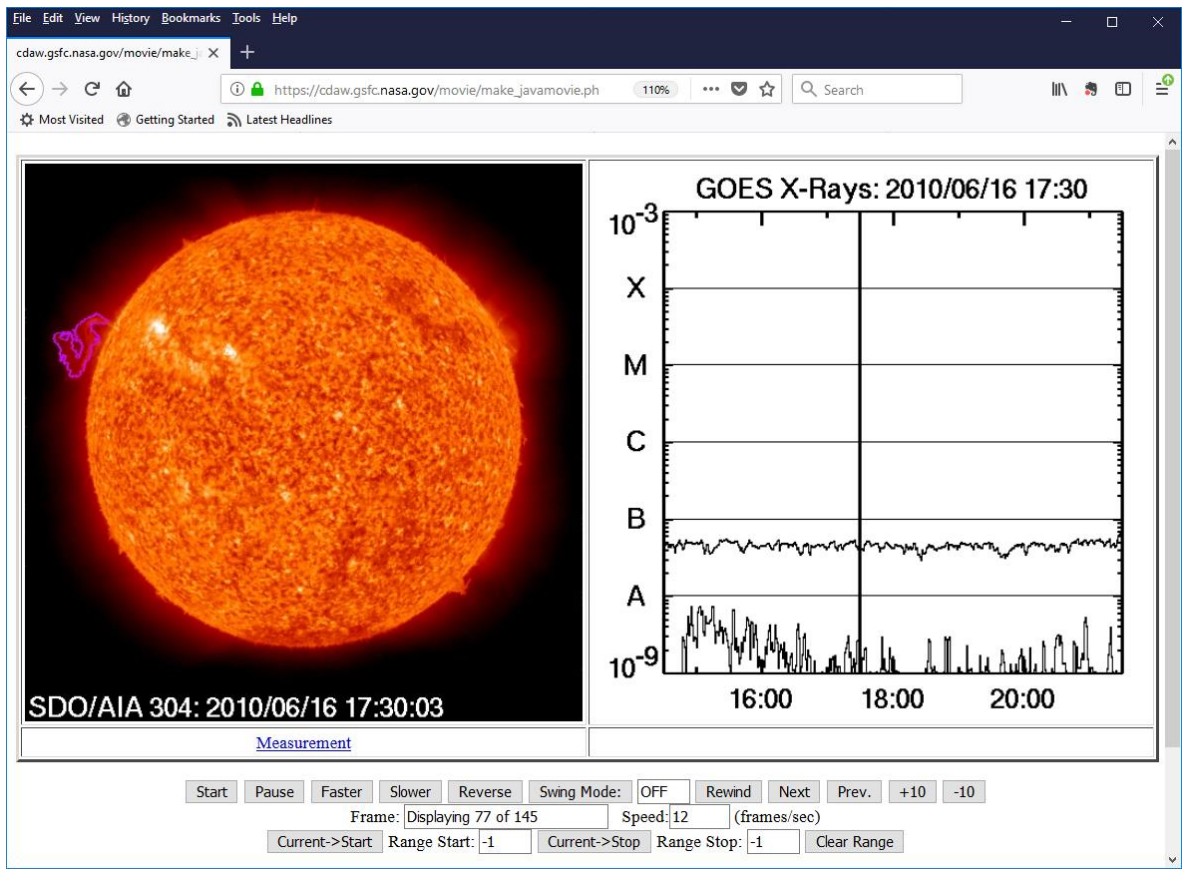

*Figure 4. One frame of the JavaScript movie for the 2010 June 16 PE. (left) SDO/AIA image with the PE above the NW limb. (right) The corresponding soft X-ray lightcurves in the GOES 1-8 Å (upper) and 0.5-4 Å (lower) channels. The vertical line in the plot corresponds to the time of the SDO/AIA frame on the left. When the movie plays, the line shifts to the right at the cadence of the AIA images.*

The list of PEs and surges automatically detected and checked for their reality are made available on line (see Fig. 3): https://cdaw.gsfc.nasa.gov/CME_list/autope/. The date and time of the prominence eruption are shown in the first and second column. The third column is central position angle (CPA), which is measured counterclockwise from the solar north in degree. The apparent latitude (LAT; 4th column) is computed as 90°-CPA for the eastern limb events and CPA-270° for western limb events. The angular width (WD; 5th column) is the angular extent of the prominence in degree. CPA, LAT, and WD are determined when the apparent area of the prominence is the maximum. The 6th column shows the source location of the prominence is the heliographic coordinate. The source location was manually determined by looking at the movies. The location of the prominence before the eruption is given. Under "Movie", links are provided to JavaScript movies that combine SDO/AIA images with GOES data (AIA304+GXRAY) and with SOHO/LASCO C2 images (+LASCO).

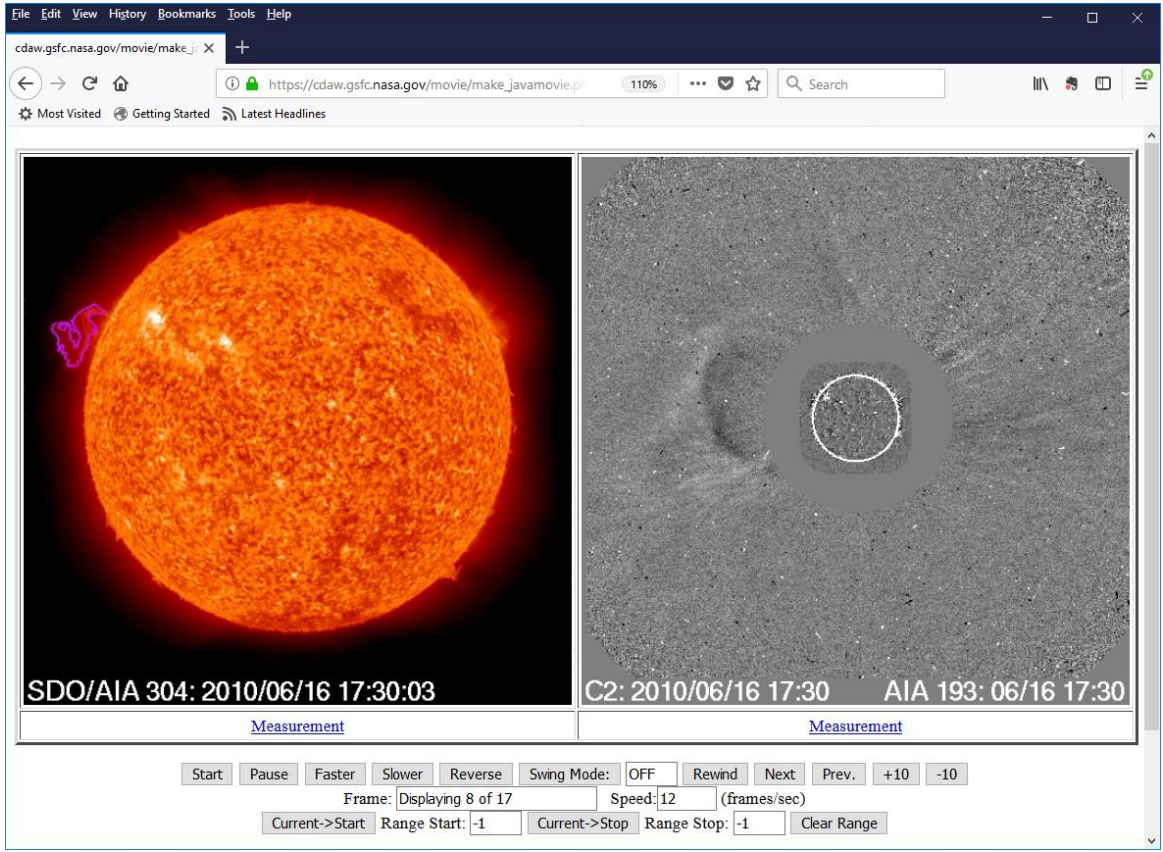

*Figure 5. One frame of the JavaScript movie for the 2010 June 16 PE. (left) SDO/AIA image with the PE above the northeast limb. (right) The corresponding SOHO/LASCO difference image with the SDO/AIA 193 Å image superposed. Note that there was a CME overlying the PE above the northeast limb. There are "Measurement" links provided under both images, which the users can use to track the height of PE/CME features and measure other parameters such as the size of the prominences.*

The JavaScript movies are useful in identifying the associated CMEs since most of the PEs leaving the Sun have an associated CME (Gopalswamy et al., 2003). Figure 4 gives one frame of a JavaScript movie showing the 2010 June 16 PE in SDO/AIA images and GOES lightcurves in the 1 – 8 Å 0.5 to 4 Å channels. In this event, there is no flare brightening associated with the PE. If the PE is on the disk, one can see a weak brightening in X-rays.

Figure 5 gives one frame of a JavaScript movie showing the 2010 June 16 PE in SDO/AIA images and the associated CME in the SOHO/LASCO FOV. We see that there is a close position-angle correspondence between the PE and the CME and that the CME is much larger than the PE. The PE is not yet seen in the LASCO frame because it is still below the occulting disk of the coronagraph. The time span of the movies can be extended to earlier or later times to track the CME-PE association and their relative positions.

## 3. Validation and Elimination of False Detections

Since artifacts are expected from automated methods, we visually examined wide events (W≥15°) to determine whether they are real events or artifacts. Out of 1,263 wide events, we found that 816 (or 64.6%) were eruptive prominences and 209 (or 16.5%) were surges. When the eruptions are seen as collimated flows from the source regions, we classify them as surges (Zirin 1988). The remaining 238 (or 18.8%) were found to be false detections.

There are two types of false detections. One is due to gradual brightness changes of the corona. If the brightness doubles within 12 hours, this algorithm recognizes it as an eruption candidate. Usually the height of such eruption candidate fluctuates, so they do not meet the condition of the 4-step monotonic increase. However, occasionally they do meet the eruption criterion. There are 90 (or 7.1%) such false detections (noise events). The other type of the false detection is due to fragmented structure of eruptions. If an eruption has more than one fragment (see, e.g., Fig. 2g,h), the algorithm may count more than one event. There are 148 (or 11.7%) such double-count events.

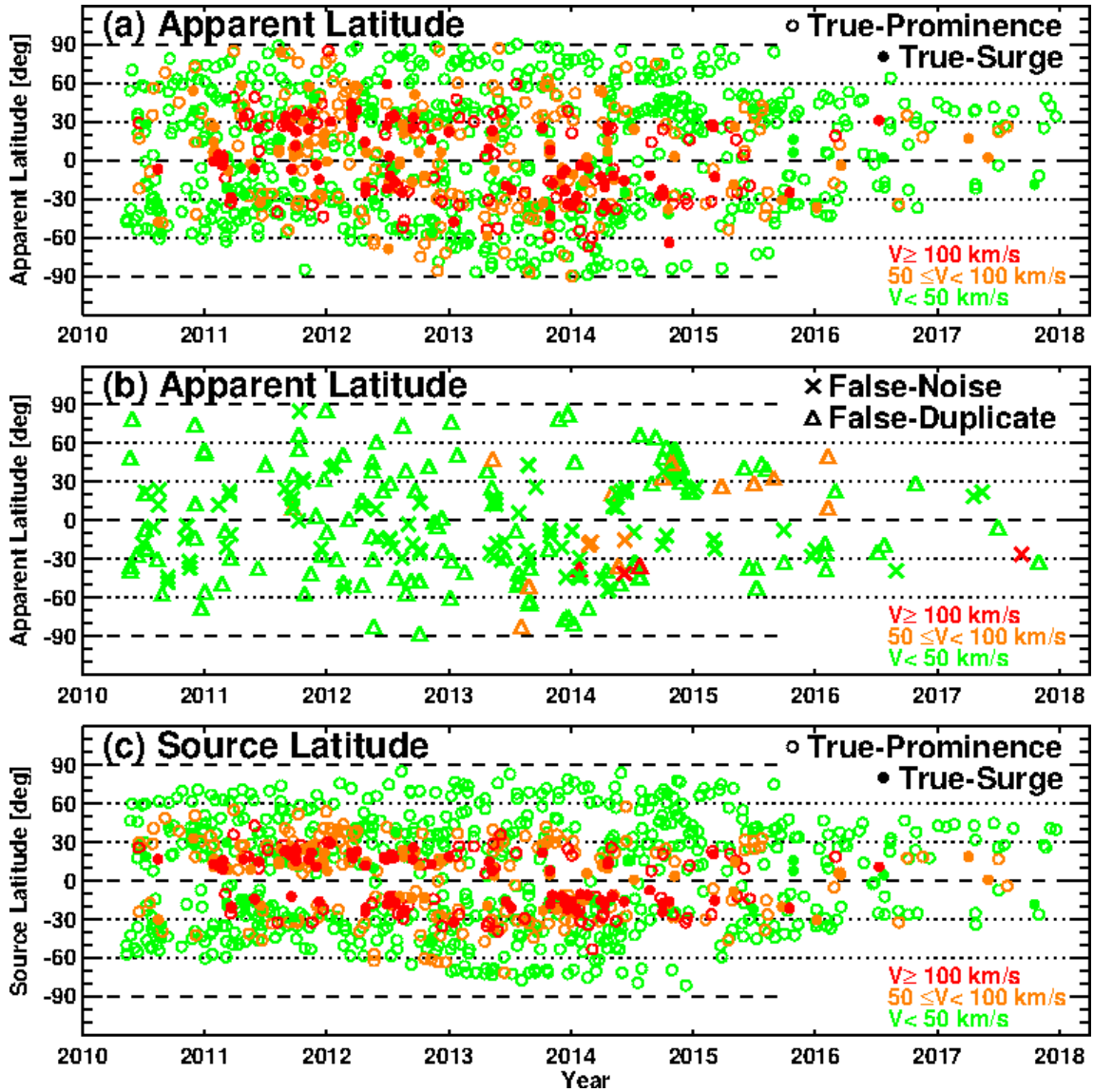

*Figure 6. (a) Solar cycle variation of apparent latitude of true prominences (open circles) and true surges (filled circles). The red, orange, and green symbols indicate fast, intermediate, and slow eruptions. (b) The latitudes of falsely-detected eruptions. The false-noise detections (crosses) are due to the brightness change of active regions, so most of them are located at low latitudes. The false-duplicates (triangles) are due to double-count of true prominences, so they are spread at all latitudes. (c) Latitudes of the source regions of the prominences (open circles) and surges (filled circles). Source latitude is the latitude at which eruptions occur.*

Tightening the condition of the monotonic height increase can reduce the false detections. If we use the 6-step monotonic-height-increase condition instead of 4-step one, the algorithm detects 60 (or 5.3%) noise events, 109 (or 9.6%) double-count events, 192 (or 16.9%) surges, and 772 (or 68.1%) prominence eruptions. The number of the false detection reduces from 238 to 169, but the algorithm fails to detect 61 true eruptions (17 surges and 44 PEs). The tightening of the condition might provide a better performance score for the automated algorithm when used unaided on its own. However, we prefer to keep the less stringent condition and visually inspect the detections in order to include as many true eruptions as possible.

Figure 6a shows the apparent latitudes of the true wide eruptions (W$\geq$15°). The open circle is for prominences and filled circle is for surges. The apparent latitude is the latitude of the centroid when the eruption area is at its maximum. The different colors indicate different speeds of the prominences: red is for fast events (V$\geq$100 km/s), orange is for intermediate-speed events (50$\leq$V$<$100 km/s) and green is for slow events (V$<$50 km/s). How to obtain the eruption speeds is described in the next Section.

Figure 6b shows the apparent latitudes of false-detection events. Most of the noise detections (crosses) appear at low latitudes, which correspond to the locations of active regions. The latitude distribution of the duplicate detections (triangles) is the same as the true prominences because the events themselves are not false. When the high-latitude prominences disappear after the polar field reversal, there are no duplicate eruptions at high latitudes.

Because the algorithm detects activity above the limb, large eruptions originating from the disk could also be detected if they extend above the limb. Because of the projection effects, the apparent latitude of such events could be very different from their latitude before the eruption (source latitude). For example, a filament located at N29E49 on 2014 April 30 and started to rise at around 9:00 UT. The filament reached the NE limb at 10:18 UT and the algorithm detected its rising motion above the limb. Because the centroid was located at *PA* of 40° at 10:28 UT, the detected apparent latitude was N50°. The offset between the source latitude and detected apparent latitude was as large as 21°.

Figure 6c shows the source latitude of the prominences and the surges. For surges the source position is measured at the foot point. It is clear that fast eruptions originate from the low and middle latitudes. The separation between northern and southern eruptions becomes clearer compared to Figure 6a. All high-latitude wide eruptions are prominence eruptions. There are no high-latitude wide surges (W≥15°) because large surges are known to be flare-related (Westin 1969, Zirin 1988).

We also examined randomly selected 500 non-wide eruptions (W<15°) and found that the false detection rate is high as 30%. Different filtering algorithms are needed to reduce the false detections for non-wide eruptions. Machine-Learning technique will help us to create good filters, but that task is beyond the scope of this paper.

**3.1 Missing Events**

We have compared our event list with the AIA Filament Eruption Catalog (McCauley et al., 2015; hereafter the FE Catalog). Out of 1014 events in the FE Catalog, 37 events don't have the event onset. We exclude them from the comparison. Out of 977 events in the FE Catalog, only 590 (or 60%) are in our list even though we have 7,368 events in the same interval (from May 2010 to October 2014). Most of them were dark filament eruptions from the disk. Our method rarely catches the dark filament eruptions only when they become bright above the limb. Other missing events meet the brightness criteria but did not meet the criteria of the monotopic height increase.

**4. Statistical Properties**

The method described detected 7,362 PEs from 2010 May 13 to 2018 December 31. The basic parameters of the eruptions are angular width, eruption speed, and location. The angular width and the location are determined when the area of an eruption reaches maximum in polar coordinate images. The extent in the PA dimension is taken as the apparent width of the eruptions. In this section, we present some results on the statistical properties of the listed PEs. While the angular size (width) and location are listed in the catalog, the speed is not. The speed can be measured by the user using the "Measurement" link provided under the images. Here we illustrate the speed measurement for an example event.

There are several heights to represent the eruption height. The leading edge height (or the maximum height) of the eruption is the most common one. Blue pluses in Figure 7 denote the maximum height of the prominence as a function of time. However, in the automated detection, the position of the leading edge is unstable when the prominence does not have clear contrast with respect to the background. Red triangles show the centroid height as a function of time. In order to obtain the prominence speed as a function of time, a series of the eruption speeds is derived from different subsets of the height-time data points. We choose 10 successive height-time points as a subset. If a prominence has *n* height-time points, we have *n-9* subsets (speeds). Blue and red curves in Fig. 7 show the erupting speed of the leading edge and the centroid, respectively of the 2011 May 30 PE shown in Fig. 2. The prominence speed increases monotonically, and then decreases when the leading edge goes out of the AIA FOV. The maximum speed is 139 km/s and 88 km/s for the leading edge and the centroid, respectively. A problem may arise on the centroid speed when a prominence erupts partially. If a part of the prominence does not erupt or falls down, the speed derived from the centroid could be unrealistically low. In order to minimize this disadvantage, we introduce the eighty-percentile height (P80 height; green diamonds) which is defined as follows. First, we make an array of the apparent height from each pixel. The maximum and the median heights are the maximum and the median values of the array, respectively. From the array, we also calculate the eighty-percentile value and use it as the P80 height. Because the maximum and median values correspond 100th and 50th percentiles, the P80 heights locate between the maximum and the median heights. The P80 height-time curve is smooth compared to that of the leading edge. The maximum P80 speed is 115 km/s, which is 17% lower than the leading edge speed. We use the P80 speed as the representative speed of prominences.

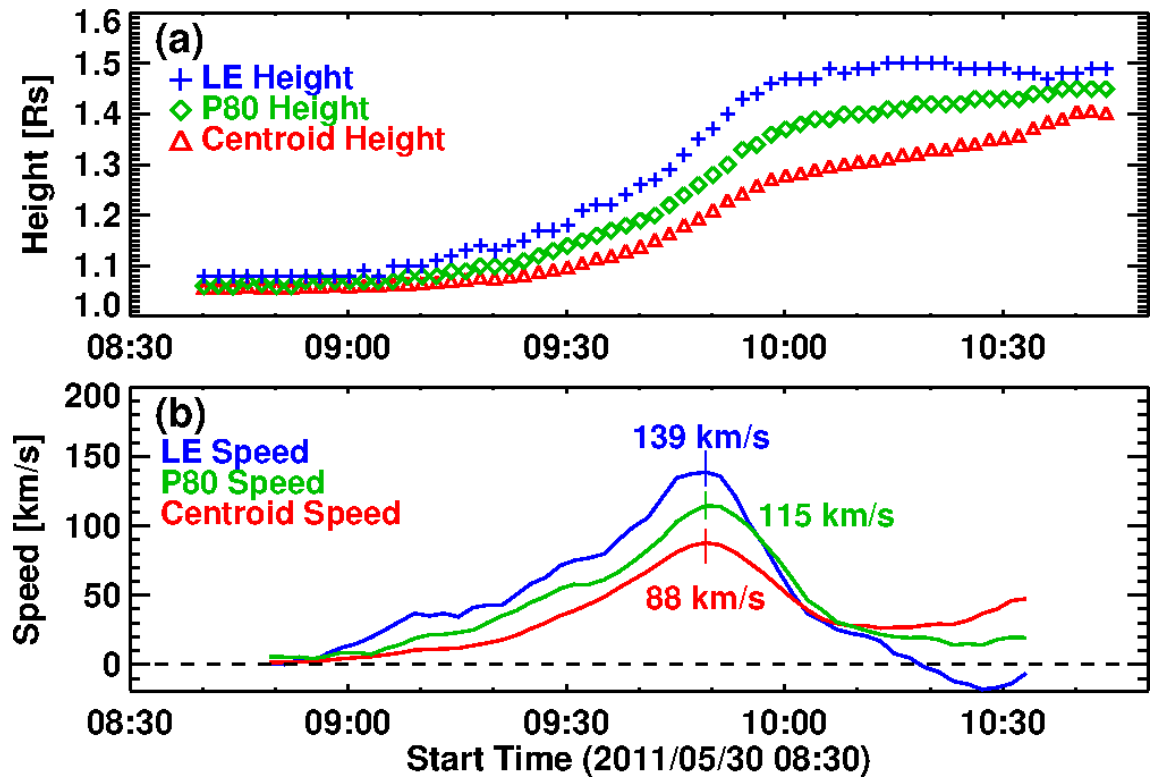

*Figure 7. (a) Height-time measurements of the 2011 May 30 PE. The height-time plot of the prominence leading edge (blue crosses), the 80-percentile height (green diamonds), and the centroid (red triangles). (b) The prominence speed derived from successive heights.*

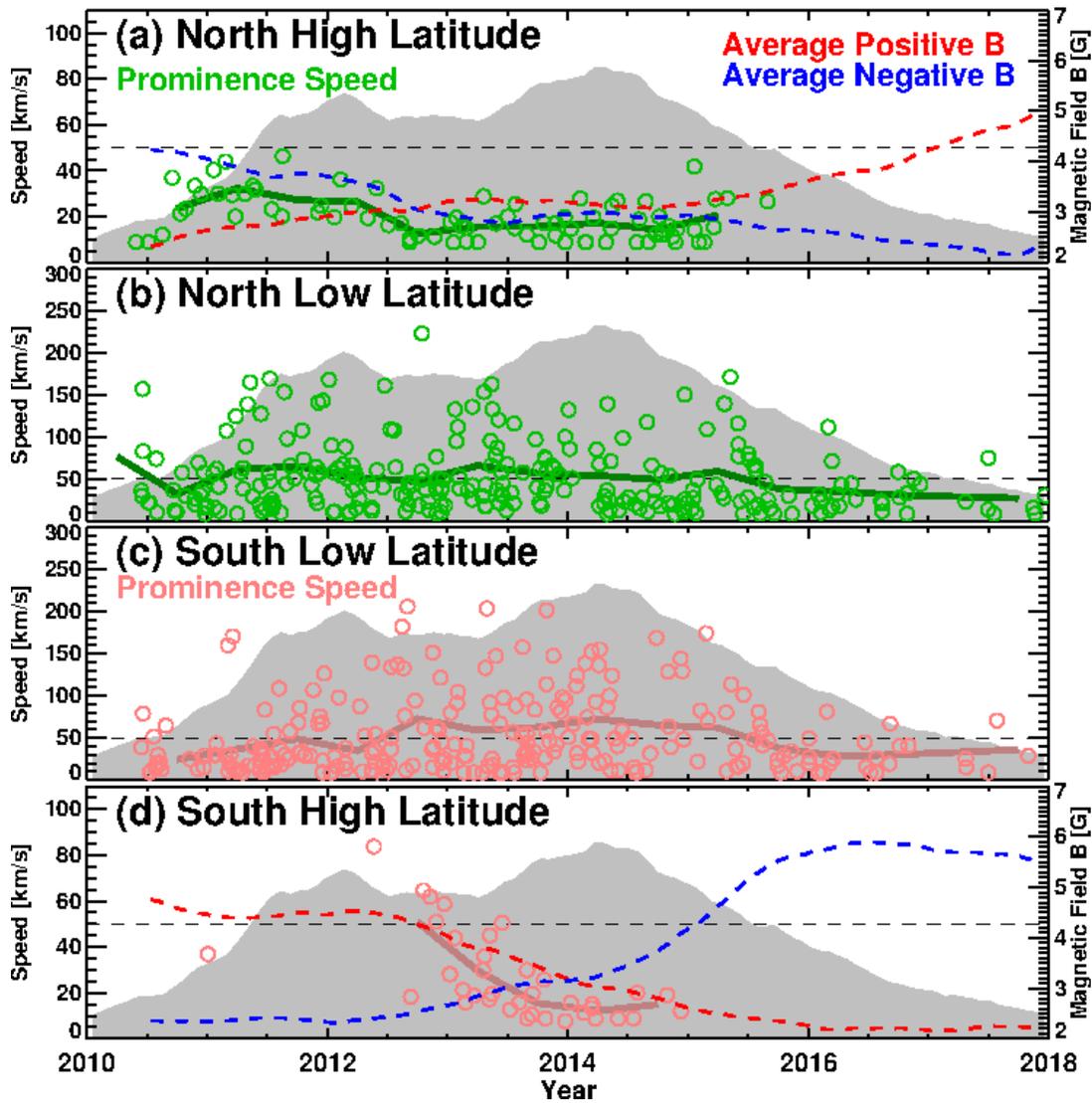

*Figure 8 – Solar cycle variation of the prominence speed of north high (a), north low (b), south low (c), and south high latitudes (d). The high and low latitude are defined as in the range of 60°-90° and 0°-40°, respectively. The average of positive (red) and unsigned negative (blue) magnetic field strength in the high latitude measured by HMI are shown (right scale). The gray background is the sunspot numbers.*

## 4. 1 Solar Cycle Variation of Prominence Speeds

Figure 8a shows the solar cycle variation of the speed of the wide eruptive prominences (W≥15°) at the northern high latitude (60°-90°). True surges and the false detections are excluded from the analysis. The gray background distribution shows the sunspot number as a reference of the solar cycle. The average speeds calculated semi-annually are shown by the solid lines. Except for the first three points in panel (a), the speed of northern high-latitude prominences in 2010 and 2011 was approximately 30 km/s with the range of 15-40 km/s. During 2012, the average speed of the high-latitude erupting prominences started to decrease and reached 18 km/s in 2013.

The dashed lines in red and blue show the average strength of positive and negative magnetic field measured by SDO/HMI in the northern high-latitude regions (60°-90°), respectively. The average positive (negative) magnetic field is the average of positive (negative) pixels in the HMI Carrington map (http://jsoc.stanford.edu/HMI/LOS_Synoptic_charts.html). In order to reduce the annual variation due to the B0 angle change, the 13-rotation smoothing is applied. The average magnetic field strength in the northern polar region was almost zero during 2012 October 8 - 2014 December 08 (Sun et al., 2015; Gopalswamy et al., 2016). The slow speeds of the northern high-latitude erupting prominences after 2013 are caused by a low magnetic field strength.

Except for an outlier on 2011 January 2, the high-latitude eruptions in the south started on 2012 May 21 (Fig. 8d). The average speed was 83.8 km/s. Similar to the northern eruptions, the speed of the high-latitude eruptions decreased with the decrease of the average positive flux. Because southern polar magnetic field reversed suddenly (within one Carrington Rotation; Gopalswamy et al. 2016), the period of slow PEs was short (~1.5 years) compared to that in the north (~3 years).

Figures 8b and 8c shows the solar cycle variation of speeds of low-latitude (0°-40°) PEs in northern and southern hemisphere, respectively. The speeds ranged up to 250 km/s. The average speeds of PEs follows solar activity in a same way as do the average speeds of CMEs (Yashiro et al. 2004; Gopalswamy et al., 2009).

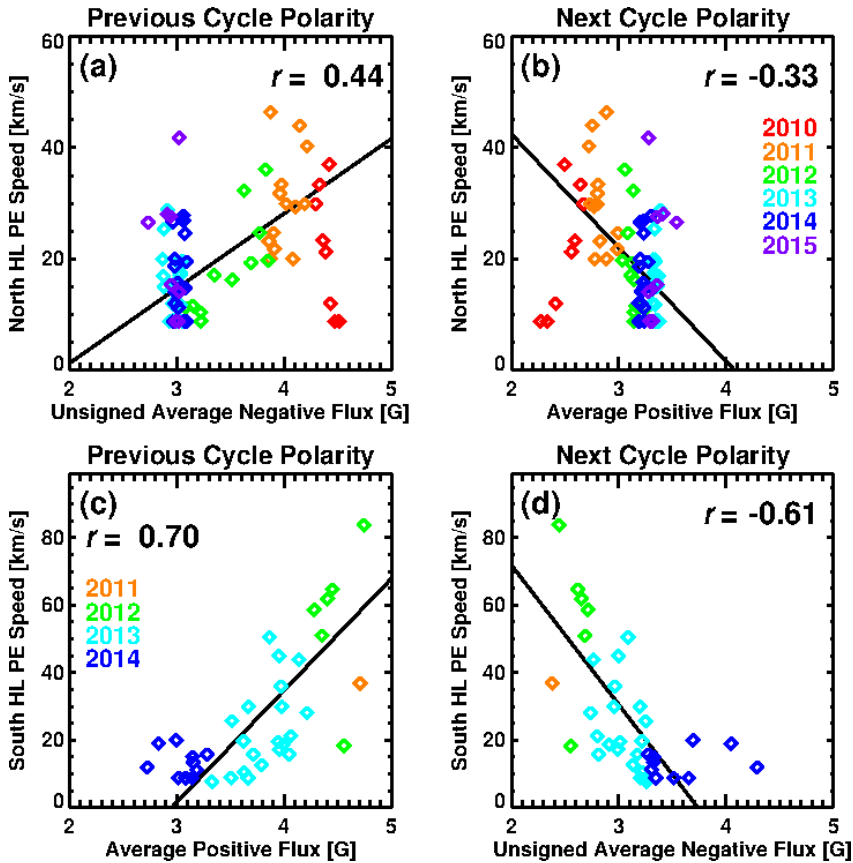

*Figure 9 – Relationship between unsigned average magnetic field strength and high latitude (HL; 60°-90°) PE speed. The top (bottom) panels are for northern (southern) high latitude events, and the left (right) panels are for positive (negative) magnetic flux, respectively. The colors indicate the year of events (see legend). The correlation coefficients (r) are shown in the each panel.*

The average PE speed in the high latitude decreased during the polarity reversal. During the polarity reversal, the magnetic field strength of the previous cycle decreases and one of the next cycle increases. In order to check which polarity is more important to determine the PE speeds, we plot the high latitude PE speeds as a function of the average magnetic field strengths of previous cycle (Fig. 9a) and the next cycle (Fig. 9b). The unsigned average flux is used for the negative polarity. The colors indicate the observed years. The northern high-latitude PE speeds decreased with decreasing of the previous (or increasing of the next) cycle polarity. The correlation coefficients (r) are 0.44 for the previous cycle and -0.33 for the next cycle. The correlation coefficients are improved to (0.58 for the previous cycle polarity and -0.52 for the next cycle polarity) if we exclude the PEs in 2010 (red points).

Figures 9c and 9d are the same but for the southern high-latitude regions. Again, the southern high-latitude PE speeds decreased with decreasing of the previous (or increasing of the next) cycle polarity. The correlation coefficients are 0.70 and -0.61 for the previous and next cycle polarity, respectively. The higher correlation was found for the southern high-latitude PEs.

In both northern and southern hemisphere, the correlation of high-latitude PE speeds and average magnetic field strength of the previous cycle polarity was higher than that of next cycle polarity. This indicates that the magnetic energy deposited at the neutral lines are controlled by the previous cycle (awaiting) polarity rather than next cycle (incoming) polarity.

## 5. Summary

We described the procedure used to identify eruptive prominences from the SDO/AIA images at 304 Å. The identified PEs are listed in an online catalog describing the central position angle, width, and heliographic coordinates of the source location. For each identified PE, we provide JavaScript movies of the PE, the associated soft X-ray lightcurve, and the SOHO/LASCO images showing the CMEs that enclose the PEs. The JavaScript movies include a tool to make measurements in the images. For example, a particular feature in the eruptive prominence can be tracked by clicking on the feature. Time and position (in units of solar radii from the disk center as well as pixel units) will be listed for each clicked point on the page. This feature enables users to measure the speed and acceleration of the eruptive prominences and those of the associated CMEs from the LASCO images.

We also analyzed the statistical properties of the detected prominences from the time of the SDO launch to the end of 2017. The results are summarized as follows:

1. The largest number of eruptive prominences were found in the active region belt. While prominences are found in all latitudes, the surges generally occur only in the active region belt.

2. The width distribution follows an exponential law with a characteristic width of ~7 degrees.

3. The centroid speed distribution also shows an exponential decrease with a characteristic speed of ~40 km/s

4. The speed of the polar eruptive prominences varies with the solar cycle.

5. Shortly before the time of the polarity reversal, the speed of the prominences becomes very small. This is because the polar magnetic field strength also becomes very small at this time. Weaker field strengths imply smaller amount of energy available to power the eruptions.

**Acknowledgments**: We thank the referees for their useful and practical comments. This research benefited from NASA's open data policy in making the SOHO, STEREO, and SDO data available on line. This work was supported by NASA's LWS TR&T program.